\documentclass[aps,prl,twocolumn,superscriptaddress]{revtex4-2}

\usepackage{graphicx}% Include figure files
\usepackage{dcolumn}% Align table columns on decimal point
\usepackage{bm}% bold math
\usepackage{rotating}

\usepackage{bm,color}

\usepackage{tabularx}
\usepackage{epstopdf}

\usepackage{ulem}
\usepackage{hyperref}
\usepackage{tikz-feynman}

\usepackage{multirow}

\usepackage{overpic}
\usepackage{verbatim}
\usepackage{amsmath}

\usepackage{lineno}
\newcommand{\gev}{\rm{GeV}}

\newcommand{\BESIIIorcid}[1]{\href{https://orcid.org/#1}{\hspace*{0.1em}\raisebox{-0.45ex}{\includegraphics[width=1em]{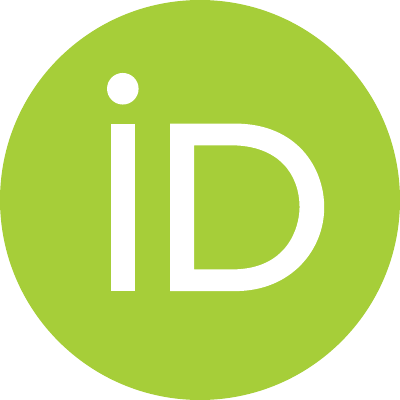}}}}

\begin{document}

\title{\boldmath
Cross sections measurement of $e^+e^-\to \Xi(1530)^0\bar\Xi^0 + c.c.$ and search for $\psi(3770)\to\Xi(1530)^0\bar\Xi^0 + c.c.$
}

%% Saved at => 2025-07-11
\author{%%
\begin{small}
\begin{center}
M.~Ablikim$^{1}$\BESIIIorcid{0000-0002-3935-619X},
M.~N.~Achasov$^{4,b}$\BESIIIorcid{0000-0002-9400-8622},
P.~Adlarson$^{79}$\BESIIIorcid{0000-0001-6280-3851},
X.~C.~Ai$^{84}$\BESIIIorcid{0000-0003-3856-2415},
R.~Aliberti$^{37}$\BESIIIorcid{0000-0003-3500-4012},
A.~Amoroso$^{78A,78C}$\BESIIIorcid{0000-0002-3095-8610},
Q.~An$^{75,61,\dagger}$,
Y.~Bai$^{60}$\BESIIIorcid{0000-0001-6593-5665},
O.~Bakina$^{38}$\BESIIIorcid{0009-0005-0719-7461},
Y.~Ban$^{48,g}$\BESIIIorcid{0000-0002-1912-0374}
% ...
\\
\vspace{0.2cm}
(BESIII Collaboration)\\
\vspace{0.2cm} {\it
$^{1}$ Institute of High Energy Physics, Beijing 100049, People's Republic of China\\
% ...
$^{4}$ Budker Institute of Nuclear Physics SB RAS (BINP), Novosibirsk 630090, Russia\\
% ...
$^{37}$ Johannes Gutenberg University of Mainz, Johann-Joachim-Becher-Weg 45, D-55099 Mainz, Germany\\
$^{38}$ Joint Institute for Nuclear Research, 141980 Dubna, Moscow region, Russia\\
% ...
$^{48}$ Peking University, Beijing 100871, People's Republic of China\\
% ...
$^{60}$ Southeast University, Nanjing 211100, People's Republic of China\\
$^{61}$ State Key Laboratory of Particle Detection and Electronics, Beijing 100049, Hefei 230026, People's Republic of China\\
% ...
$^{75}$ University of Science and Technology of China, Hefei 230026, People's Republic of China\\
% ...
$^{78}$ University of Turin and INFN, (A)University of Turin, I-10125, Turin, Italy; (B)University of Eastern Piedmont, I-15121, Alessandria, Italy; (C)INFN, I-10125, Turin, Italy\\
$^{79}$ Uppsala University, Box 516, SE-75120 Uppsala, Sweden\\
% ...
$^{84}$ Zhengzhou University, Zhengzhou 450001, People's Republic of China\\
\vspace{0.2cm}
$^{\dagger}$ Deceased\\
% ...
$^{b}$ Also at the Novosibirsk State University, Novosibirsk, 630090, Russia\\
% ...
$^{g}$ Also at State Key Laboratory of Nuclear Physics and Technology, Peking University, Beijing 100871, People's Republic of China\\
% ...
}
\end{center}
\vspace{0.4cm}
\end{small}
}
%% ends here %%

\begin{abstract}
Using $e^+e^-$ collision data  collected with the BESIII detector corresponding to an integrated luminosity of 44.2 fb$^{-1}$, we measure the Born cross sections for the process $e^+e^- \to \Xi(1530)^{0} \bar{\Xi}^{0} + c.c.$ at forty-eight center-of-mass energies between 3.51 and 4.95 GeV. The potential signal from non-$D\bar{D}$ decays for $\psi(3770)$, i.e. $\psi(3770)\to \Xi(1530)^{0} \bar{\Xi}^{0}+ c.c.$, is investigated by fitting the dressed cross section, no obvious signal is found. The upper limit of $\mathcal{B}(\psi(3770) \to \Xi(1530)^{0} \bar{\Xi}^{0}$) at 90\% confidence level is given.  

\end{abstract}

%\keywords{}

\maketitle


%\section{Introduction}

%\clearpage
The first charmonium resonance above $D\bar{D}$ production threshold, $\psi$(3770), was observed in the 1970s~\cite{3773}. 
It decays mostly to $D\bar{D}$ final states as the OZI rule dictates, while the fractions of hadronic and radiative transitions to lower lying charmonium states and decays into light hadrons are small~\cite{nondd1the1, nondd1the2}. However, the BES experiment measured the branching fraction for $\psi(3770) \to$ non-$D\bar{D}$ to be $14.7\pm3.2$\%~\cite{nondd0,nondd1,nondd3,nondd4}, in contrast to the CLEO experiment that reported a branching fraction of $(-3.3\pm1.4^{+6.6}_{-4.8})$\%~\cite{cleo}.
The first observation of a non-$D\bar{D}$ decay channel, $\psi(3770) \to \pi^+\pi^- J/\psi$~\cite{decay1}, was reported by BES experiment in 2005. Subsequently, the CLEO collaboration observed additional exclusive non-$D\bar{D}$ decays of $\psi(3770)$, including hadronic  ($\pi^0\pi^0 J/\psi$ and $\eta J/\psi$~~\cite{decay2}) and radiative transitions ($\gamma\chi_{c0}$ and $\gamma\chi_{c1}$~\cite{decay3,decay4}). Recently, the BESIII collaboration reported a first measurement of the branching fraction for the inclusive decay $\psi(3770)\to J/\psi X$ ($X=\pi^{+}\pi^{-}$, $\pi^{0}\pi^{0}$, $\eta$, $\pi^{0}$, $\gamma\gamma$) of $(0.5\pm0.2)$\% ~\cite{new3770}. This result is compatible with the sum of the related branching fractions~\cite{pdg} for $\psi(3770)\to\pi^{+}\pi^{-}J/\psi$, $\pi^{0}\pi^{0}J/\psi$, $\eta J/\psi$, $\gamma\chi_{cJ}$ with $J=0,1,2$.

Besides the study of the hadronic and radiative transitions to lower lying charmonium states discussed above, are the decays $\psi(3770)$ $\to$ $B\bar{B}$ ($B$ = baryon) of interest. This is due to the simple topology of the final states and a relatively clean decay mechanism that is assumed to be dominated by a three-gluon or one-photon process. Up to now, the experimental information on the decays of $\psi(3770)$ into baryon pairs is scarce above the open-charm threshold. Experimental studies~\cite{prestudy1, prestudy2, prestudy3, prestudy4, prestudy6, prestudy7,prestudy8} of the processes $e^+ e^- \to$ $B\bar{B}$  were performed from $\sqrt{s} = 3.51 $ to $ 4.95 $ GeV by the BESIII and Belle experiments. Except for the evidences for $\psi(3770)\to \Lambda\bar\Lambda$~\cite{prestudy6} and $\psi(3770)\to$ $\Xi^-\bar\Xi^+$~\cite{prestudy7}, there is no evidence for $\psi(3770)$ decaying to octet-decuplet baryon pairs ($B_{10} \bar{B}_8$)~\cite{pdg}. 
%{\new{Previous papers have studied $J/\psi$ decays to  $B_8 \bar{B}_{10}$}}~\cite{b8b10,b8b102}, and
Based on the SU(3)-flavor symmetry, the decays $\psi \to$ $B\bar{B}$ into baryon pairs of the same multiplet are allowed, but the decays $\psi \to$ $B_{10} \bar{B}_8$ are forbidden~\cite{forbidden, forbidden1, forbidden2}. Nevertheless, violations of these predictions have long been known. For instance, $J/\psi \to \Xi^-\bar{\Xi}(1530)^+$, $J/\psi\to\bar{\Sigma}^+\Sigma(1835)^-$, $J/\psi\to\bar{\Sigma}^-\Sigma(1835)^+$, 
$\psi(3686)\to\Xi(1530)^+\bar\Xi^-$, $\psi(3686)\to\Xi(1530)^0\bar\Xi^0$, etc., were reported by the DM2 and BESIII Collaborations~\cite{dm2, jilaoshi,xiongfei,charged,jilaoshi2}.
Measurement of cross sections of the $e^+ e^- \to$ $B_{10} \bar{B}_8$ final states above the
open charm threshold and investigations of the $\psi(3770)\to$ $B_{10} \bar{B}_8$ decay will therefore provide additional information to understand the nature of $\psi(3770)$.

This Letter reports the measurement of Born cross section for 
the $e^+e^- \to \Xi(1530)^{0} \bar{\Xi}^{0}$ with $\Xi(1530)\to\Xi^{-}\pi^{+}$ and $\bar{\Xi}^{0}\to \bar{\Lambda}\pi^{0}$ (charge conjugation is implied throughout this letter) reaction, at center-of-mass (CM) energies ($\sqrt{s}$) between 3.51 and 4.95 GeV. The measurement uses $e^{+}e^{-}$ collision data corresponding to a total integrated luminosity of $44.2$ fb$^{-1}$~\cite{lum1,lum2} (the values of integrated luminosity at each energy point are available in the Supplemental material~\cite{supple}), collected by the BESIII detector~\cite{bes3} at the BEPCII collider~\cite{bepc2}. In addition, we also search for $\psi(3770)$ in the line-shape of the dressed cross section.

The detection efficiency is established by Monte Carlo (MC) simulations using a sample of 100,000 events for each CM energy point, with a phase space (PHSP) model by the {\sc kkmc}~\cite{kkmc1,kkmc2} generator including effects of the beam energy spread and ISR corrections. The $e^+e^-\to\Xi(1530)^0\bar\Xi^0$ process and its subsequent decays are simulated with the PHSP model by the {\sc evtgen}~\cite{even1,even2} generator. The BESIII geometric description and the detector response are modeled with a Geant4-based~\cite{geant4} software package. The cylindrical core of the BESIII detector covers 93\% of the full solid angle and consists of a helium-based multilayer drift chamber (MDC), a plastic scintillator time-of-flight system (TOF), and a CsI(Tl) electromagnetic calorimeter (EMC), which are all enclosed in a superconducting solenoidal magnet providing a 1.0 T magnetic field. This analysis employs a single-baryon tagging technique in which we tag the $\Xi(1530)^0$ candidate through the decay chain $\Xi(1530)^0 \to \Xi^- \pi^+$, with $\Xi^- \to \Lambda \pi^-$ and $\Lambda \to p \pi^-$, while the $\bar{\Xi^0}$ partner is extracted by the recoil mass spectrum of the tagged $\Xi(1530)^{0}$ signal.

\begin{figure}[htbp]
\begin{center}
\includegraphics[width=0.45\textwidth]{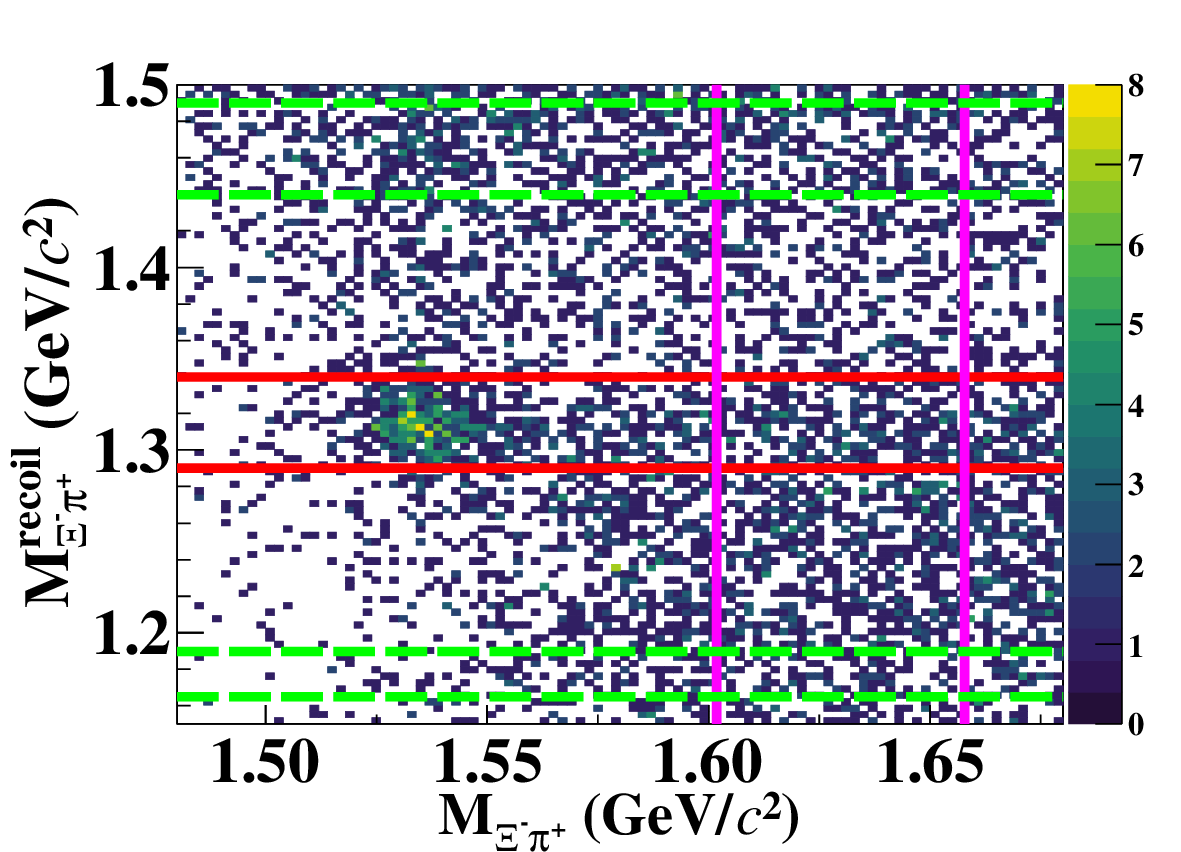}
\end{center}
\caption{The distribution of $M_{\Xi^- \pi^+}^{\rm recoil}$ versus $M_{\Xi^- \pi^+}$ of the accepted candidates in data at $\sqrt{s} = $ 3.773 GeV. The red lines represent the $\bar{\Xi}^0$ signal region, the dashed-green lines indicate the $\bar{\Xi}^0$ sideband regions, and the pink lines represent the $\Xi(1530)^0$ sideband region.}
\label{fig:scatter2d}
\end{figure}

Charged tracks detected in the MDC are required to be within a polar angle $\theta$ range of $|cos\theta|$ \textless~0.93, where $\theta$ is defined with respect to the $\boldmath{z}$-axis, which is the symmetry axis of the MDC. Particle identification (PID) for charged tracks combines measurements of the energy deposited in the MDC $dE/dx$ and the flight time in the TOF to form likelihoods $\mathcal{L}_{h}$ ($h = p, K, \pi$) for each hadron $h$ hypothesis. Tracks are identified as protons (pions) when the proton (pion) hypothesis has the greatest likelihood, $i.e.$ $\mathcal{L}_{p(\pi)}$ \textgreater~$\mathcal{L}_{K}$ and $\mathcal{L}_{p(\pi)}$ \textgreater~$\mathcal{L}_{\pi(p)}$.  The $\Lambda$ candidates are reconstructed from $p\pi^-$ pairs with an invariant mass within $\pm 7$ MeV/$c^2$ of the $\Lambda$ nominal mass~\cite{pdg}.  A secondary-vertex fit~\cite{sec-vertexfit} is performed with all the $p\pi^-$ combinations; those with $\chi^2$ \textless ~500 are kept for further analysis. To further suppress background events, the decay length of the $\Lambda$ is required to be positive. In the case of multiple candidates, the one with an unconstrained mass closest to the nominal mass is retained. The $\Xi^-$ candidates are reconstructed from $\Lambda\pi^-$ pairs with an invariant mass within $\pm 6.5$ MeV/$c^2$ of the $\Xi^-$ nominal mass~\cite{pdg}. A secondary-vertex fit is performed with all the $\Lambda\pi^-$ combinations. The candidate with an invariant mass closest to the nominal mass is retained if there is more than one candidate in an event. The decay length of $\Xi^-$ is required to be positive to further suppress background. The $\Xi(1530)^0$ candidates are reconstructed by minimizing $|M_{\Xi^- \pi^+} - M_{\Xi(1530)^0}|$, where the $M_{\Xi^- \pi^+}$ is the invariant mass of $\Xi^-\pi^+$ and $M_{\Xi(1530)^0}$ is the nominal $\Xi(1530)^0$ mass~\cite{pdg} (the method has been validated and will not cause bias for the $\Xi(1530)^0$ signal peak). The mass window of $\Xi(1530)^0$ is optimized to be $|M_{\Xi^- \pi^+} - M_{\Xi(1530)^0}| ~\textless$ 13 MeV/$c^2$. Furthermore, the $\bar{\Xi}^0$ mass window is chosen as $|M_{\Xi^- \pi^+}^{\rm recoil} - M_{\bar{\Xi}^0}|~ 
\textless$ 25 MeV/$c^2$, where $M^{\rm recoil}_{\Xi^-\pi^+}$ is the recoil mass of $\Xi^-\pi^+$ and $M_{\bar{\Xi}^0}$ is the nominal $\bar{\Xi}^0$ mass~\cite{pdg}. These mass windows are chosen by optimizing the figure-of-merit (FOM = $S/\sqrt{S+B}$), where $S$ is the number of signal MC events and $B$ is the number of estimated background events from the inclusive MC sample.
Figure~\ref{fig:scatter2d} shows the distribution of $M_{\Xi^- \pi^+}^{\rm recoil}$ versus $M_{\Xi^- \pi^+}$ of the accepted candidates in the data at $\sqrt{s}=3.773~\gev$. The red-horizontal lines denote the $\bar{\Xi}^0$ signal region, the dashed-black and pink lines denote the $\bar{\Xi}^0$ and $\Xi(1530)^0$ sideband regions, respectively.

\begin{figure*}[htbp]
\begin{center}
\includegraphics[width=0.3\textwidth]{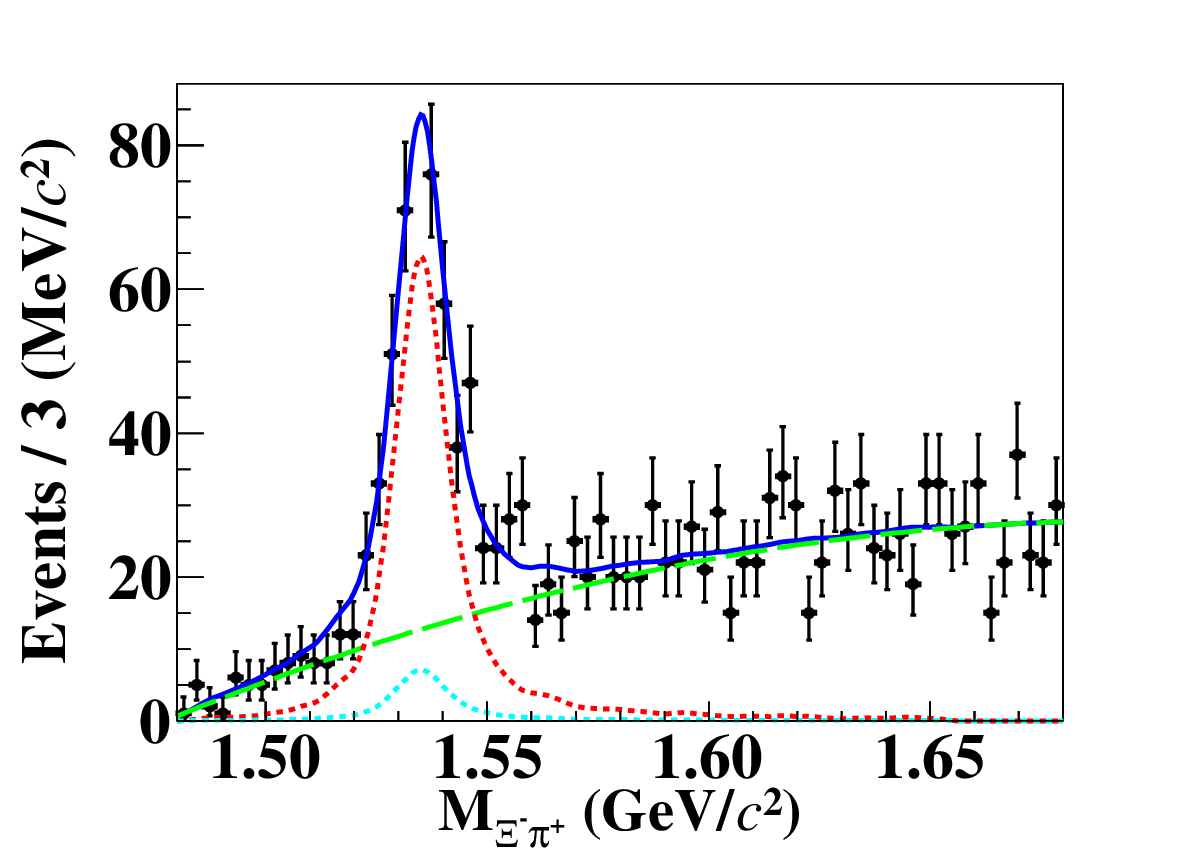}
\put(-75, -6){\bf \scalebox{0.75}{(a)}}
\includegraphics[width=0.3\textwidth]{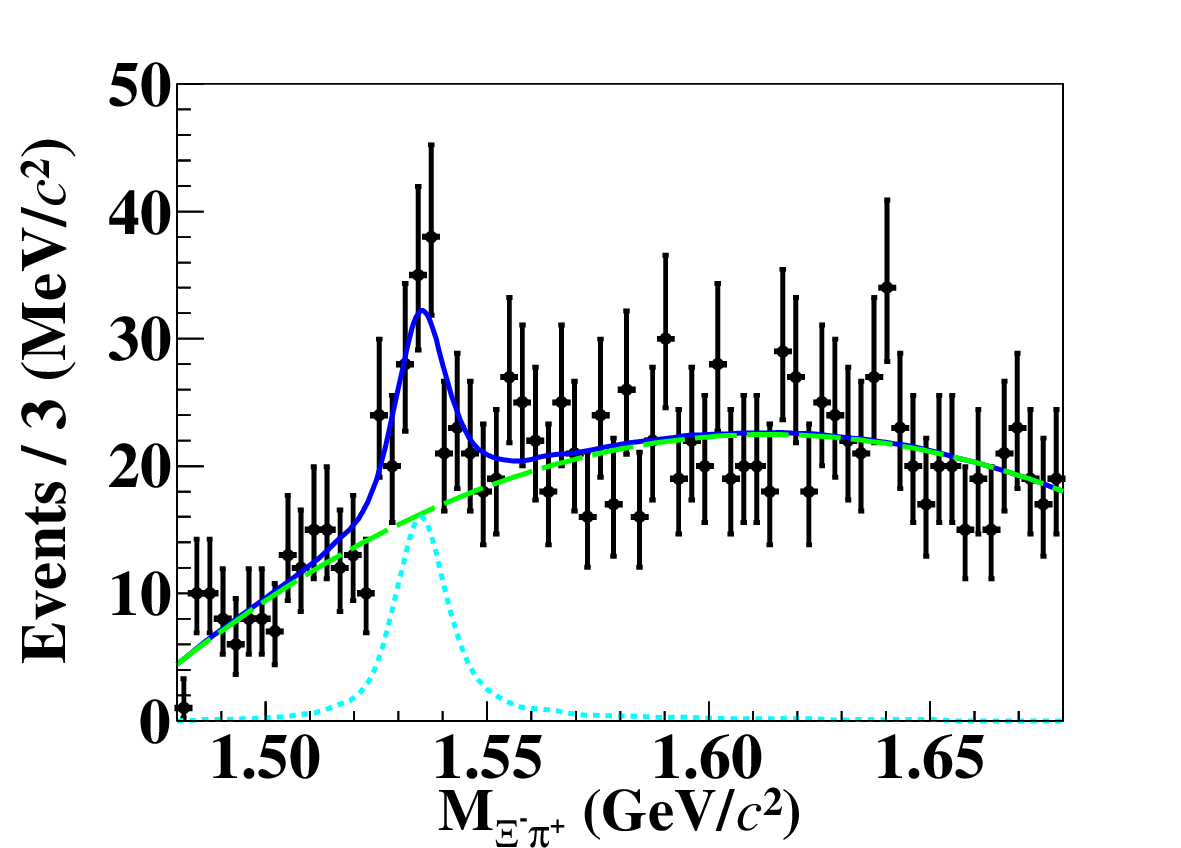}
\put(-75, -6){\bf \scalebox{0.75}{(b)}}
\includegraphics[width=0.3\textwidth]{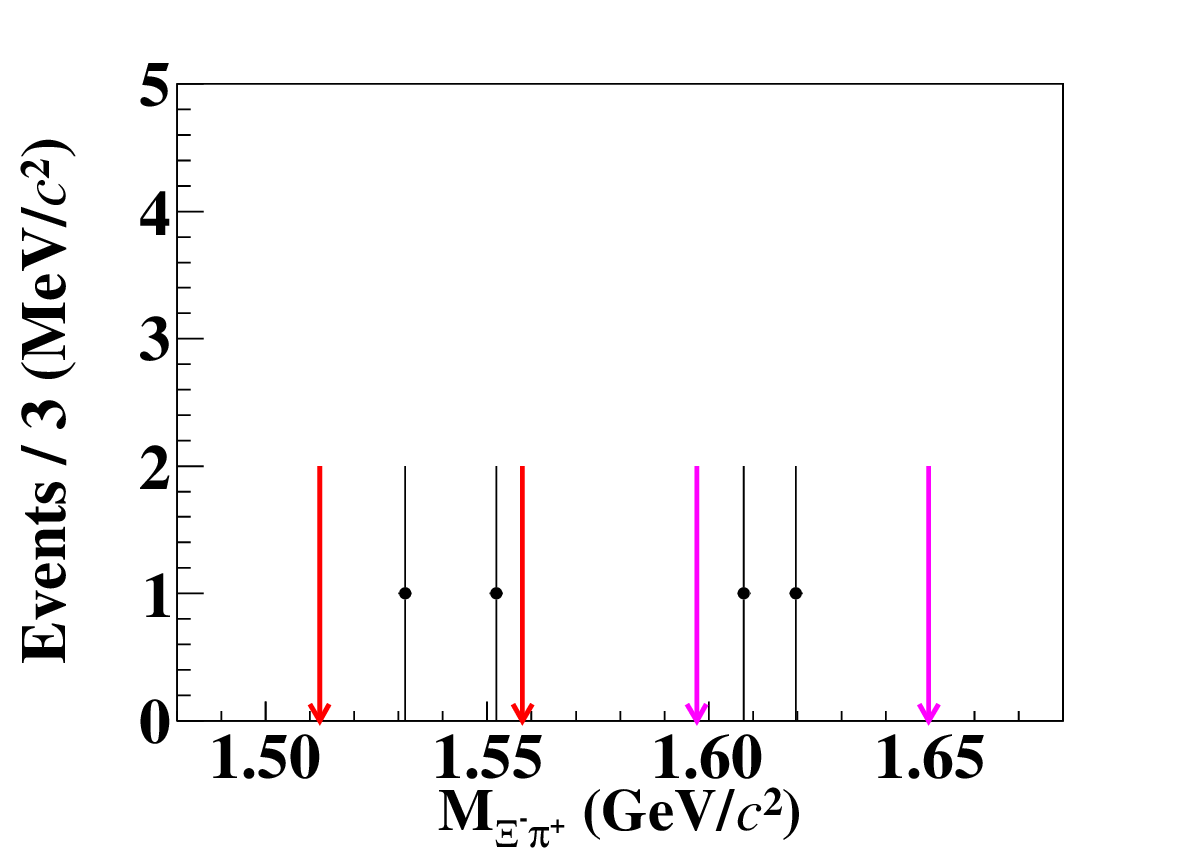}
\put(-75, -6){\bf \scalebox{0.75}{(c)}}
\end{center}
\setlength{\abovecaptionskip}{0.7pt}
\caption{The simultaneous fit to the $\boldmath{M}_{\Xi^-\pi^+}$ distributions in the $\bar{\Xi}^0$ signal (a) and sideband (b) regions for data at $\sqrt{s}= $ 3.773 GeV. The dashed-green lines indicate the Chebyshev function describing the background shape of the $\bar{\Xi}^0$ signal and sideband regions. The dashed-cyan lines indicate the peaking background of $\Xi(1530)^0$ in both $\bar{\Xi}^0$ signal and sideband regions. The dashed-red lines indicate the signal shape of $\Xi(1530)^0$. (c) The $M_{\Xi^- \pi^+}$ distribution at $\sqrt{s}= $ 4.246 GeV, where the red and pink arrows indicate the $\Xi(1530)^0$ signal and sideband regions, respectively.}
 \label{fig:scatter}
\end{figure*}

Two methods are used in this analysis (fitting and counting) to determine the signal yield, depending on the statistics at different energy points.
The signal yield at $\sqrt{s} = $ 3.773 GeV is obtained by performing
a simultaneous fit on the $M_{\Xi^-\pi^+}$ distributions with events in the $\bar{\Xi}^0$ signal and sideband regions. Events in the $\bar{\Xi}^0$ sideband regions are normalized with a scale factor, $R_1=0.44$, which is obtained by comparing the number of smooth background events in $\bar{\Xi}^0$ signal and sideband regions, and then fixed in the simultaneous fit. The $\bar{\Xi}^0$ signal region is determined to be [1.277, 1.352] GeV/$c^2$, and the $\bar{\Xi}^0$ sideband regions are [1.440, 1.490] and [1.165, 1.190] GeV/$c^2$. Figure~\ref{fig:scatter} (b) and (c) show the results of the simultaneous fit at $\sqrt{s}=$ 3.773 GeV, where the MC-simulated signal shapes are shared for the $\bar{\Xi}^0$ signal and sideband regions in the simultaneous fit.

Due to the low statistics, the signal yield at the other energy points is extracted counting the events as
\begin{equation}
N^{\rm obs} = N^{S} - R_{2} \cdot N^{\rm bkg},
\end{equation}
where $R_{2}$ is the scale factor related to the $\Xi(1530)^0$ signal and sideband regions in the $M_{\Xi^-\pi^+}$ distribution, determined to 0.52 by using a similar method as for $R_{1}$ with the data at $\sqrt{s}=$ 3.773 GeV. The $\Xi(1530)^0$ signal and sideband regions are chosen as [1.511, 1.560] GeV/$c^2$ and [1.602, 1.658] GeV/$c^2$, respectively; $N^{S}$ and $N^{\rm bkg}$ are the numbers of events in the above two corresponding regions, obtained by counting as shown in Fig.~\ref{fig:scatter} (d).  Limited by the low statistics, the $R_{2}$ values at other CM energy points are assumed to be the same as the one at $\sqrt{s}=$ 3.773 GeV. The statistical significance is estimated by $\int_{-s}^s 1/\sqrt{2\pi}e^{-x^2/2} dx =1-P(N_s)$, where $P(N_s)=1-\Sigma_{n=0}^{N_s-1}(b^n/n!)e^{-b}$ is the experimental P-value for each energy point, $s$ and $b$ is the number of signal and background events in the signal region. The uncertainty of $N^{\rm obs}$ and its upper limit are computed by the TRolke method~\cite{troke}.

The Born cross section (BCS) $\sigma^{B}$ of the $e^+e^- \to \Xi(1530)^{0} \bar{\Xi}^{0}$ process is given by 
\begin{equation}\label{func:dressed cs}
\begin{split}
\sigma^{B} = \frac{N^{\rm obs}}{\mathcal{L}\cdot(1+\delta)\cdot \frac{1}{|1-\Pi|^{2}}\cdot \epsilon \cdot \mathcal{B}},
\end{split}
\end{equation}
where $\mathcal{L}$ is the integrated luminosity, $(1+\delta)$ is the ISR correction factor, $\frac{1}{|1-\Pi|^{2}}$ is the vacuum polarization (VP) correction factor, $\epsilon$ is the selection efficiency, and $\mathcal{B}$ is the product of the branching fractions of $\Xi(1530)^0\to\Xi^-\pi^+$, $\Xi^-\to\Lambda\pi^-$ and $\Lambda\to p\pi^-$. The VP correction factor is calculated according to Ref.~\cite{vp}. The ISR correction factor is determined with an iterative weighting method~\cite{iter}. 

\begin{figure}[htbp]
\includegraphics[width=0.43\textwidth]{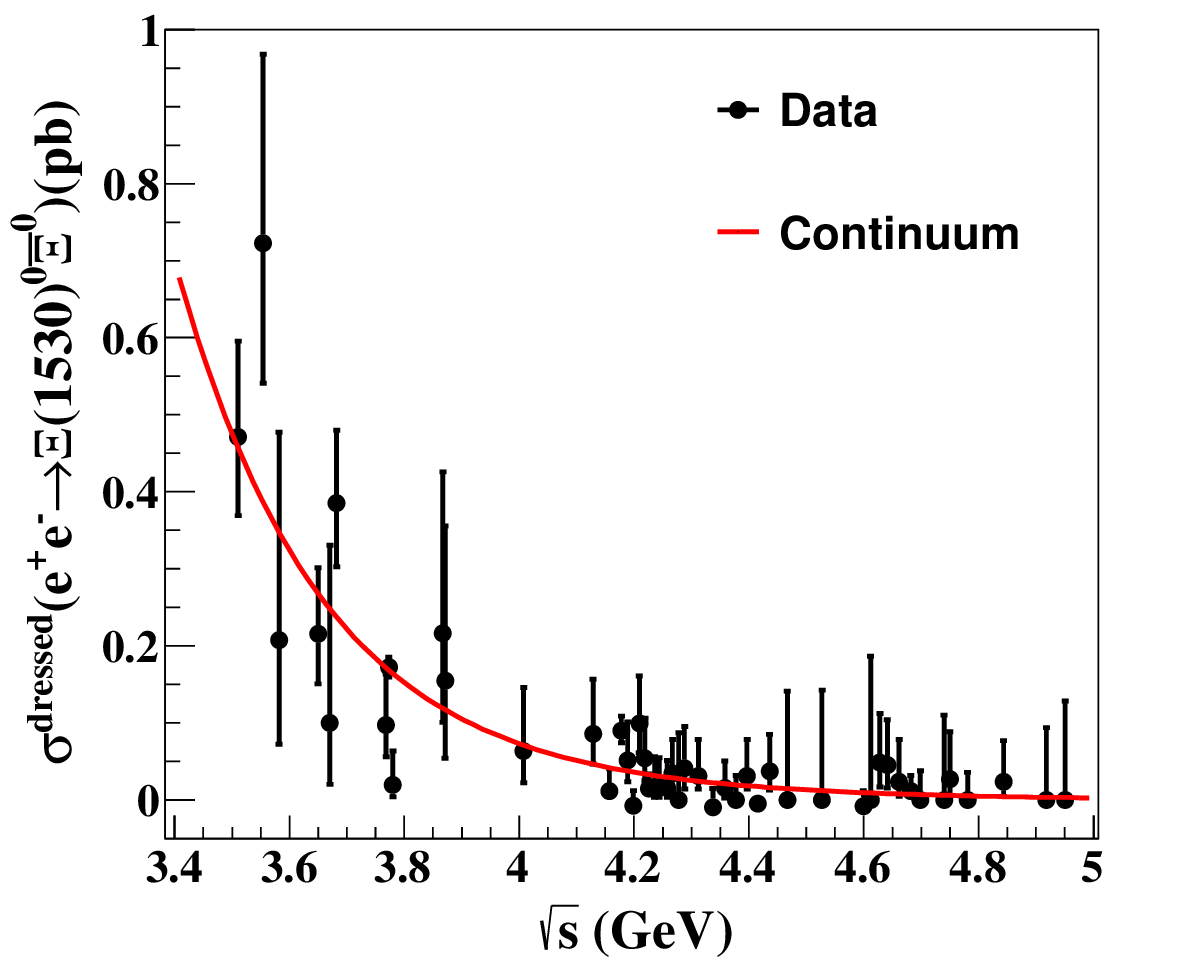}

\includegraphics[width=0.43\textwidth]{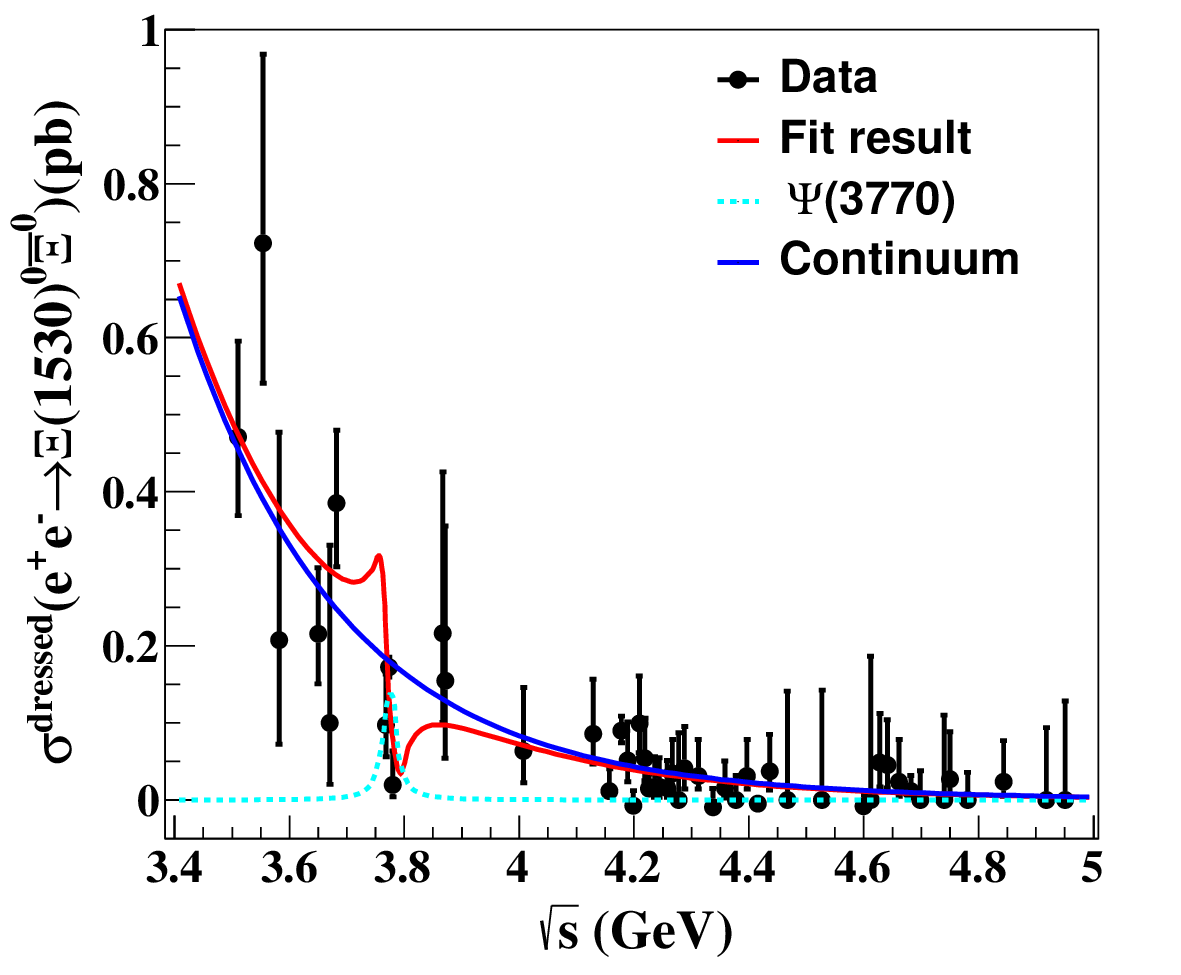}

\setlength{\abovecaptionskip}{0.72pt}
\caption{Fits to the dressed cross section of $e^+e^- \to \Xi(1530)^{0} \bar{\Xi}^{0}$ with the assumption of a PL function (upper plot) and a PL function plus a $\psi(3770)$ resonance (lower plot). The dots with error bars are the dressed cross sections, the red solid lines are the fit results.} 
\label{fig:cx}
\end{figure}

To study the possible decay $\psi(3770) \to \Xi(1530)^{0} \bar{\Xi}^{0}$, a maximum-likelihood fit is performed to the obtained dressed cross section ($\sigma^{\rm dressed}$=$\sigma^{B} \cdot \frac{1}{|1-\Pi|^{2}}$). The baseline model $\sigma^{\rm dressed}(\sqrt{s})$
is written as
\begin{equation}
|A\cdot \sqrt{p(\sqrt{s})}\cdot \frac{1}{\sqrt{s}^{N}} +
e^{i\phi}BW(\sqrt{s};m,\Gamma,\Gamma_{e^{+}e^{-}} \mathcal{B})\cdot \sqrt{\frac{p(\sqrt{s})}{p(M)}}|^{2}.
\end{equation} 
Here, the first term is a power-law (PL) function,  in the second term $BW$ is used to describe the resonance, defined as
\begin{equation}
BW(\sqrt{s}) = \frac{\sqrt{12\pi\Gamma_{e^{+}e^{-}}\mathcal{B}(R\to \Xi(1530)^{0}\bar\Xi^{0})\Gamma}}{s-\boldmath{M}^{2}+i\boldmath{M}\Gamma}, 
\end{equation}
$p(\sqrt{s})$ is the two-body PHSP factor, $A$ and $N$ are parameters of the PL function.
In the fit, $A$, $N$, the relative phase $\phi$, and the product of the electronic partial width and the branching fraction of the resonance decaying into the $\Xi(1530)^{0} \bar{\Xi}^{0}$ final state ($\Gamma_{ee}\mathcal{B}(R\to \Xi(1530)^{0}\bar\Xi^{0})$) are free parameters. The mass $M$ and the total width $\Gamma$ are fixed to the $\psi(3770)$ resonance parameters from the PDG~\cite{pdg}. The upper plot in Figure~\ref{fig:cx}  shows the results of the fits to the dressed cross sections with only the PL function, whereas the lower plot shows the result for adding an extra assumed $\psi(3770)$ amplitude. The fit parameters with only the PL function are $A=24.4\pm22.6, N=8.0\pm0.7$.
Table~\ref{table:solution} shows the parameters with an additional assumed $\psi(3770)$ amplitude for the two solutions. Here, the possible solutions are evaluated based on a two-dimensional scan method, which scans all the pairs of $\Gamma_{e^{+}e^{-}}\mathcal{B}$ and $\phi$ in the parameter space. The $-$ln$\cal{L}$ distribution as a function of the production branching fraction multiplied by the electronic partial width and the relative phase $\phi$ are shown in the Supplemental Material~\cite{supple}. The significance of the assumed $\psi(3770)$ resonance is calculated by comparing the likelihood values with and without involving the $\psi(3770)$ resonance in the fit and taking the change of the number of degrees of freedom into account. No obvious signal is found,  the $\mathcal{B}(\psi(3770) \to \Xi(1530)^0 \bar{\Xi}^0)$ is determined to be $6.91 \times 10^{-5}$  at 90\% confidence level including systematic uncertainty by using the Bayesian method~\cite{upper}.  

\begin{table}[htbp]
\centering
\caption{The fitted resonance parameters for $\Gamma_{ee}\mathcal{B}$ and $\phi$ (rad) with the two solutions. The errors includes both statistical and systematic uncertainties.}\label{table:solution}
\begin{center}

\begin{tabular}{ccc}
\hline\hline
$\psi(3770)$ &$\Gamma_{ee}\mathcal{B}$($10^{-3}$ eV)  & $\phi(rad)$ \\
\hline
Solution I &$13.3\pm6.8 (< 18.8)$ &$-2.0\pm0.4$ \\
Solution II &$3.7\pm6.2$ &$-2.6\pm0.4$ \\

\hline
\end{tabular}
\end{center}
\end{table}

The sources of the systematic uncertainties in the BCS measurement are detailed below.

The integrated luminosities of each energy point are measured using the Bhabha events, which have uncertainties of about 1.0\% below 4.0 GeV~\cite{sys lum1}, 0.7\% from 4.0 to 4.6 GeV~\cite{sys lum2}, and 0.5\% above 4.6 GeV~\cite{sys lum3}.

The uncertainties caused by the $\pi^{+}$ tracking [coming from the $\Xi(1530)^0$ mother particle] and  PID are investigated with a control sample of $\psi(3686) \to  \pi^{+}\pi^{-}p\bar{p}$ events. The observed difference in the efficiencies between data and MC simulation, 1.0\%, is assigned as the uncertainty for the tracking or PID. 
The uncertainty of the $\Xi^-$ reconstruction is estimated to be 5.1\% with a control sample of $\psi(3686)\to\Xi^-\bar\Xi^+$~\cite{5.1} which including the efficiency of the
 tracking/PID, and $\Lambda$ reconstruction, the decay length of $\Lambda/\Xi$ and mass window of $\Lambda/\Xi$. 
%The uncertainty associated with the efficiency for the decay length and mass window requirements on $\Lambda$  candidates is estimated with a control sample of $J/\psi \to \Xi^{-}\bar{\Xi}^{+}$~\cite{xixi}, and the difference in selection efficiency between the data and MC simulation, 0.1\%, and 0.3\%, respectively, are taken as the uncertainty. The case for $\Xi^-$ candidates is estimated using the same method and gives 0.5\% and 1.3\%, respectively.
The uncertainty due to the $\Xi(1530)^0$ mass window is estimated by fitting the distributions of $M_{\Xi^-\pi^+}$ from data and MC simulation The
difference in the accepted efficiencies between data and MC simulation, 3.6\%, is taken as the uncertainty.
It is worth noting that this uncertainty is only valid for the data samples with low statistics, but not for $\sqrt{s}$ = 3.773 GeV, whose signal yield is obtained from the simultaneous fit.
The uncertainty due to the $\bar\Xi^0$ mass window is estimated by fitting the distributions of $M_{\Xi^-\pi^+}^{\rm recoil}$ from data and MC simulation. The difference in the accepted efficiencies between data and MC simulation, 2.3\%, is taken as the uncertainty. The MC models for the signal channel require the angular-distribution parameters of the baryon pairs as input. The uncertainty due to the angular distribution of the $e^+e^-\to\Xi(1530)^0\bar{\Xi}^0$ process is investigated by changing the model from PHSP to J2BB3~\cite{even1,even2}. The resulting differences of 3.7\% for $\Xi(1530)^0$ and 4.9\% for $\bar\Xi(1530)^0$ are taken as the uncertainties.

%in determining the signal yield are calculated from the fit results with all the parameters fixed to the best values. 
The uncertainty due to the scale factors is caused by the fit parameters, and it is estimated by generating multi-dimensional Gaussian random numbers using the covariance matrix values from the fit as input. The standard deviation of the resultant number of $\Xi(1530)^0$ events, 1.1\%,  is taken as the systematic uncertainty.
The uncertainty introduced by the wrong combination of the $\pi^+$ from the $\Xi(1530)^0$ decay is assessed by comparing the signal yields with and without the corresponding component included in the fit. The difference in the cross section, 2.1\%, is taken as the uncertainty.

The uncertainty of the input cross section line shape includes two parts: (i) the uncertainty of the input line shape of the cross sections, (ii) the uncertainty of the resonance parameters For part (i), the input line shape can be varied by changing the parameters, which are used in the fit function, as random numbers satisfying Gaussian distributions. Here, we use the line shape after four iterations as the nominal value, we sample it 200 times according to its error matrix, and perform a Gaussian fit on the sampling results. The width of the Gaussian function is taken as the corresponding systematic uncertainty. For part (ii), the uncertainty is estimated by varying the mass and width of $\psi(3770)$ by $\pm 1\sigma$ with respect to the PDG values. Then, the  difference of the $\mathcal{B}(\psi(3770) \to \Xi(1530)^0 \bar{\Xi}^0)$ with the nomimal value is taken as the systematic uncertainty. The total systematic uncertainty due to the input line shape is the quadratic sum of these two parts. Their values are given in the Supplemental Material~\cite{supple}.

The uncertainties due to the quoted branching fractions of the intermediate decays $\Xi(1530)^0\to\Xi^-\pi^+$, $\Xi^-\to\Lambda\pi^-$ and $\Lambda\to p\pi^-$, are 3.7\%, 0.0\% and 0.8\%~\cite{pdg}, respectively, where the 3.7\% uncertainty of the branching fraction of
$\Xi(1530)^0\to\Xi^-\pi^+$ is conservatively taken as the upper limit for the branching fraction of $\Xi(1530)\to\gamma\Xi$~\cite{pdg}.

The total systematic uncertainty in the BCS measurements at each energy point is obtained by summing the individual contributions in quadrature, under the assumption that these sources are independent, and is given in the Supplemental Material~\cite{supple}.

In conclusion, using $e^+e^-$ collision data at forty-eight center-of-mass energies between 3.51 and 4.95 GeV, the Born cross sections for the  $e^+e^- \to \Xi(1530)^{0} \bar{\Xi}^{0}$ reaction are measured. The dressed cross section of this reaction is fitted under the assumption of a $\psi(3770)$ amplitude plus a continuum contribution. No obvious signal is found, the $\mathcal{B}(\psi(3770) \to \Xi(1530)^0 \bar{\Xi}^0)$ at 90\% confidence level is given. These measurements provide important experimental information on the correlation between vector charmonium states and the $e^+e^- \to \Xi(1530)^{0} \bar{\Xi}^{0}$ production, and will be helpful to understand the nature of $\psi(3770)$.

The BESIII Collaboration thanks the staff of BEPCII (https://cstr.cn/31109.02.BEPC) and the IHEP computing center for their strong support. This work is supported in part by National Key R\&D Program of China under Contracts Nos. 2023YFA1606000, 2023YFA1606704; National Natural Science Foundation of China (NSFC) under Contracts Nos. 11635010, 11935015, 11935016, 11935018, 12025502, 12035009, 12035013, 12061131003, 12192260, 12192261, 12192262, 12192263, 12192264, 12192265, 12221005, 12225509, 12235017, 12361141819; the Chinese Academy of Sciences (CAS) Large-Scale Scientific Facility Program; the Strategic Priority Research Program of Chinese Academy of Sciences under Contract No. XDA0480600; CAS under Contract No. YSBR-101; 100 Talents Program of CAS; The Institute of Nuclear and Particle Physics (INPAC) and Shanghai Key Laboratory for Particle Physics and Cosmology; ERC under Contract No. 758462; German Research Foundation DFG under Contract No. FOR5327; Istituto Nazionale di Fisica Nucleare, Italy; Knut and Alice Wallenberg Foundation under Contracts Nos. 2021.0174, 2021.0299; Ministry of Development of Turkey under Contract No. DPT2006K-120470; National Research Foundation of Korea under Contract No. NRF-2022R1A2C1092335; National Science and Technology fund of Mongolia; Polish National Science Centre under Contract No. 2024/53/B/ST2/00975; STFC (United Kingdom); Swedish Research Council under Contract No. 2019.04595; U. S. Department of Energy under Contract No. DE-FG02-05ER41374

\bibliography{basename of .bib file}

\end{document}

\title{\boldmath Supplemental Material for  ``Measurement of cross section of $e^+e^-\to \Xi(1530)^0\bar\Xi^0 + c.c.$ and search for $\psi(3770)\to\Xi(1530)^0\bar\Xi^0 + c.c.$
"}

\maketitle

\onecolumngrid

%\clearpage
%\section{Numerical results at each data sample}

\begin{table*}[htbp]
\centering
\caption {
Numerical results for the dressed Born cross sections of $e^+e^-\to \Xi(1530)^0\bar\Xi^0$ at different energy points~\cite{cm1,cm2}. Upper limits on $N^{\rm obs}$ and  $\sigma^{B}$ at 90\% C.L. are listed in parentheses if the significances are less than 3$\sigma$. The first uncertainties are statistical and the second are systematic. The $S(\sigma)$ is  the statistical significance of the signal.}\label{table:nobs list}
\begin{center}
\scalebox{0.95}{
\begin{tabular}{ccccccccc}
\hline
$\sqrt s$ (GeV) &$\mathcal{L}$ ($pb^{-1}$)&$N^{S}$  & $N^{\rm bkg}$ & $N^{\rm obs}$ &$\frac{1}{|1-\Pi|^{2}}$ &$(1+\delta)\cdot\epsilon$ &$\sigma^{B}(fb)$ &$S(\sigma)$\\
\hline
$3.51000$ &405.7 & $29$ & $15$ & $21.2^{+5.6}_{-4.6}$ &1.045 &0.13 &$471.4^{+124.5}_{-102.3}\pm47.6$ &4.4 \\

$3.55400$ &129.4 & $12$ & $1$ & $11.5^{+3.9}_{-2.9}$ &1.041 &0.14 &$722.9^{+245.2}_{-182.3}\pm73.0$  &3.3 \\

$3.58200$ &85.7 & $3$ & $2$ & $2.0^{+2.6}_{-1.3}(\textless 5.1)$ &1.039 &0.13 &$207.2^{+269.3}_{-135.0}\pm21.0$(\textless529.5 ) &1.3 \\

$3.65000$ &410.0 & $12$ & $2$ & $11.0^{+4.4}_{-3.3}$ &1.021 &0.15 &$215.4^{+86.2}_{-64.6}\pm21.8$ &3.2\\

$3.67010$ &84.7 & $1$ & $0$ & $1.0^{+2.3}_{-0.8}$(\textless 3.7) &1.011 &0.14 &$100.1^{+230.2}_{-80.1}\pm10.1$(\textless370.3 ) &1.0 \\

$3.68200$ &404.0 & $26$ & $14$ & $18.7^{+4.6}_{-4.0}$ &1.021 &0.14 &$385.3^{+94.8}_{-82.4}\pm38.9$ &4.1 \\

$3.76800$ &415.8 & $8$ & $6$ & $5.0^{+3.4}_{-2.2}(\textless 8.2)$ &1.054 &0.14 &$97.2^{+66.1}_{-40.8}\pm9.8$(\textless159.5 ) &2.1\\

$3.77300$ &20274.8 & -      & -     & $417.4\pm31.0$         &1.056 &0.14 &$172.7^{+12.8}_{-12.8}\pm17.4$ & 20 \\

$3.78000$ &410.0 & $4$ & $6$ & $1.0^{+2.3}_{-0.8}(\textless 3.6)$ &1.059 &0.15 &$19.3^{+44.4}_{-15.5}\pm2.0$(\textless69.6 ) &1.3\\

$3.86704$ &108.9 & $3$ & $0$ & $3.0^{+2.9}_{-1.6}(\textless 6.9)$ &1.051 &0.15 &$216.5^{+209.3}_{-115.5}\pm21.9$(\textless498.0 ) &1.7 \\

$3.87131$ &110.3 & $3$  & $2$ & $2.0^{+2.6}_{-1.3}(\textless 5.1)$&1.051 &0.14 &$154.6^{+201.0}_{-100.5}\pm15.6$(\textless394.3 ) &1.3 \\

$4.00762$ &482.0 & $2$  & $0$ & $2.0^{+2.6}_{-1.3}(\textless 5.4)$&1.044 &0.12 &$63.5^{+82.5}_{-41.2}\pm6.4$(\textless171.3 ) &1.4 \\

$4.12848$ &401.5 & $4$  & $1$ & $3.5^{+2.9}_{-1.6}(\textless 7.5)$&1.052 &0.12 &$85.7^{+71.0}_{-39.2}\pm8.7$(\textless183.7 ) &1.8 \\

$4.15740$ &408.7& $2$  & $3$ &$0.5^{+1.3}_{-0.3}(\textless 2.7)$&1.051 &0.13 &$11.3^{+29.5}_{-6.8}\pm1.1$(\textless61.3 ) &0.9\\

$4.17800$ &3189.0 & $38$  & $12$ &$31.8^{+6.7}_{-5.6}$&1.054 &0.13 &$90.1^{+19.0}_{-15.9}\pm9.1$ &5.4\\

$4.18800$ &526.7 & $3$  & $0$ &$3.0^{+2.9}_{-1.6}(\textless 6.9)$&1.056 &0.13 &$51.5^{+49.8}_{-27.5}\pm5.2$(\textless118.4 ) &1.7\\

$4.19890$ &526.0 & $0$  & $1$ &$-0.5^{+1.3}_{-0.3}(\textless 1.2)$&1.056 &0.14 &$-7.7^{+20.1}_{-4.6}\pm0.8$(\textless18.5 ) &0.0\\

$4.20939$ &517.1 & $6$ & $1$ & $5.5^{+3.4}_{-2.1}(\textless 10.3)$&1.057 &0.13 &$99.4^{+61.4}_{-37.9}\pm10.0$(\textless186.1 ) &2.3\\

$4.21893$ &514.6 & $3$  & $0$ & $3.0^{+2.9}_{-1.6}(\textless 6.9)$&1.056 &0.13 &$53.9^{+52.1}_{-28.7}\pm5.4$(\textless123.9 ) &1.7 \\

$4.22630$ &1100.9 & $3$  & $2$ &$2.0^{+2.6}_{-1.3}(\textless 5.1)$&1.057 &0.14 &$14.8^{+19.3}_{-9.6}\pm1.5$(\textless37.8 ) &1.3\\

$4.23570$ &530.3 & $1$  & $0$ & $1.0^{+2.3}_{-0.8}(\textless 3.7)$&1.056 &0.13 &$17.0^{+39.1}_{-13.6}\pm1.7$(\textless62.9 ) &1.0 \\

$4.24397$ &538.1 & $2$  & $2$ & $1.0^{+2.3}_{-0.8}(\textless 3.6)$&1.056 &0.13 &$16.7^{+39.1}_{-13.3}\pm1.7$(\textless60.0 ) &1.0\\

$4.25800$ &828.4 & $1$  & $0$ &$1.0^{+2.3}_{-0.8}(\textless 3.7)$&1.054 &0.14 &$15.5^{+35.6}_{-12.4}\pm1.6$(\textless57.3 ) &1.0\\

$4.26681$ &531.1 & $3$  & $2$ & $2.0^{+2.6}_{-1.3}(\textless 5.1)$&1.053 &0.13 &$34.2^{+44.5}_{-22.2}\pm3.5$(\textless87.3 ) &1.3 \\

$4.27770$ &175.7 & $0$  & $0$ &$0.0^{+1.8}_{-0.0}(\textless 2.0)$&1.054 &0.14 &$0^{+87.3}_{-0}\pm0$(\textless97.0 ) &0.0\\

$4.28788$ &502.4 & $3$  & $2$ & $2.0^{+2.6}_{-1.3}(\textless 5.1)$&1.053 &0.11 &$41.5^{+53.9}_{-27.0}\pm4.2$(\textless105.8 ) &1.3\\

$4.31205$ &501.2 & $2$  & $1$ & $1.5^{+2.3}_{-0.8}(\textless 4.5)$&1.052 &0.11 &$31.0^{+47.6}_{-16.5}\pm3.1$(\textless93.1 ) &1.2\\

$4.33740$ &505.0 & $0$  & $1$ &$-0.5^{+1.3}_{-0.3}(\textless 1.2)$&1.051 &0.13 &$-9.1^{+23.7}_{-5.5}\pm0.9$(\textless21.9 ) &0.0\\

$4.35800$ &544.0 & $1$  & $0$ &$1.0^{+2.3}_{-0.8}(\textless 3.7)$&1.051 &0.14 &$15.4^{+35.4}_{-12.3}\pm1.6$(\textless56.9 ) &1.0\\

$4.37740$ &522.7 & $0$  & $0$ &$0.0^{+1.8}_{-0.0}(\textless 2.0)$&1.051 &0.13 &$0^{+32.1}_{-0}\pm0$(\textless35.7 ) &0.0\\

$4.39645$ &507.8 & $2$  & $1$ & $1.5^{+2.3}_{-0.8}(\textless 4.5)$&1.051 &0.11 &$31.0^{+47.5}_{-16.5}\pm3.1$(\textless92.9 ) &1.2\\

$4.41560$ &1090.7 & $1$  & $3$ &$-0.6^{+1.3}_{-0.3}(\textless 1.1)$&1.052 &0.14 &$-4.7^{+10.3}_{-2.4}\pm0.5$(\textless8.7 ) &0.5\\

$4.43624$ &569.9& $2$  & $0$ & $2.0^{+2.6}_{-1.3}(\textless 5.4)$&1.054 &0.11 &$37.0^{+48.1}_{-24.1}\pm3.7$(\textless100.0 ) &1.0 \\

$4.46710$ &111.1 & $0$  & $0$ &$0.0^{+1.8}_{-0.0}(\textless 2.0)$&1.055 &0.13 &$0^{+141.5}_{-0}\pm0$(\textless157.2 ) &0.0\\

$4.52710$ &112.1 & $0$  & $0$ &$0.0^{+1.8}_{-0.0}(\textless 2.0)$&1.055 &0.13 &0$^{+142.8}_{-0}\pm0$(\textless158.7 ) &0.0\\

$4.59950$ &586.9 & $0$  & $1$ &$-0.5^{+1.3}_{-0.3}(\textless 1.2)$&1.055 &0.13 &$-7.8^{+20.2}_{-4.7}\pm0.8$(\textless18.6 ) &0.0\\

$4.61190$ &103.7 & $0$  & $0$ &$0.0^{+1.8}_{-0.0}(\textless 2.0)$&1.055 &0.11 &$0^{+186.3}_{-0}\pm0$(\textless207.0 ) &0.0\\

$4.62800$ &521.5 & $2$  & $0$ & $2.0^{+2.6}_{-1.3}(\textless 5.3)$&1.054 &0.09 &$48.7^{+63.3}_{-31.7}\pm4.9$(\textless129.1 ) &1.0\\

$4.64091$ &551.7 & $2$  & $0$ & $2.0^{+2.6}_{-1.3}(\textless 5.4)$&1.054 &0.09 &$45.4^{+59.0}_{-29.5}\pm4.6$(\textless122.6 ) &1.0\\

$4.66124$ &529.4 & $1$  & $0$ & $1.0^{+2.3}_{-0.8}(\textless 3.7)$&1.054 &0.09 &$23.9^{+54.9}_{-19.1}\pm2.4$(\textless88.3 ) &1.0\\

$4.68192$ &1667.4& $2$  & $1$ & $1.5^{+2.6}_{-1.3}(\textless 4.5)$&1.039 &0.09 &$11.7^{+20.2}_{-10.1}\pm1.2$(\textless35.0 ) &1.2\\

$4.69880$ &535.5& $0$  & $0$ &$0.0^{+1.8}_{-0.0}(\textless 2.0)$&1.055 &0.10 &$0^{+37.7}_{-0}\pm0$(\textless41.9 ) &0.0\\

$4.73970$ &163.9& $0$  & $0$ &$0.0^{+1.8}_{-0.0}(\textless 2.0)$&1.055 &0.12 &$0^{+110.2}_{-0}\pm0$(\textless122.4 ) &0.0\\

$4.75000$ &366.6& $1$  & $0$ &$1.0^{+2.3}_{-0.8}(\textless 3.7)$&1.054 &0.12 &$26.8^{+61.7}_{-21.4}\pm2.7$(\textless99.2 ) &1.0\\

$4.78054$ &511.5& $0$  & $0$ &$0.0^{+1.8}_{-0.0}(\textless 2.0)$&1.055 &0.11 &$0^{+36.1}_{-0}\pm0$(\textless40.2 ) &0.0\\

$4.84307$ &525.2& $1$  & $0$ & $1.0^{+2.3}_{-0.8}(\textless 3.7)$&1.056 &0.10 &$23.4^{+53.7}_{-18.7}\pm2.4$(\textless86.4 ) &1.0 \\

$4.91802$ &207.8& $0$  & $0$ &$0.0^{+1.8}_{-0.0}(\textless 2.0)$&1.056 &0.11 &$0^{+92.0}_{-0}\pm0$(\textless104.5 ) &0.0\\

$4.95093$ &159.3& $0$  & $0$ &$0.0^{+1.8}_{-0.0}(\textless 2.0)$&1.056 &0.10 &$0^{+128.5}_{-0}\pm0$(\textless142.8 ) &0.0\\
\hline
\end{tabular}
}
\end{center}
\end{table*}

\newpage
%\section{The $-$Ln$\cal{L}$ distribution versus $\Gamma_{ee}{\cal{B}}(\psi(3770))\to\Xi(1530)^0\bar{\Xi}^0)$ and the relative phase $\phi$ }
\begin{figure*}[htbp]
\begin{center}
\begin{overpic}
[width=0.7\textwidth]{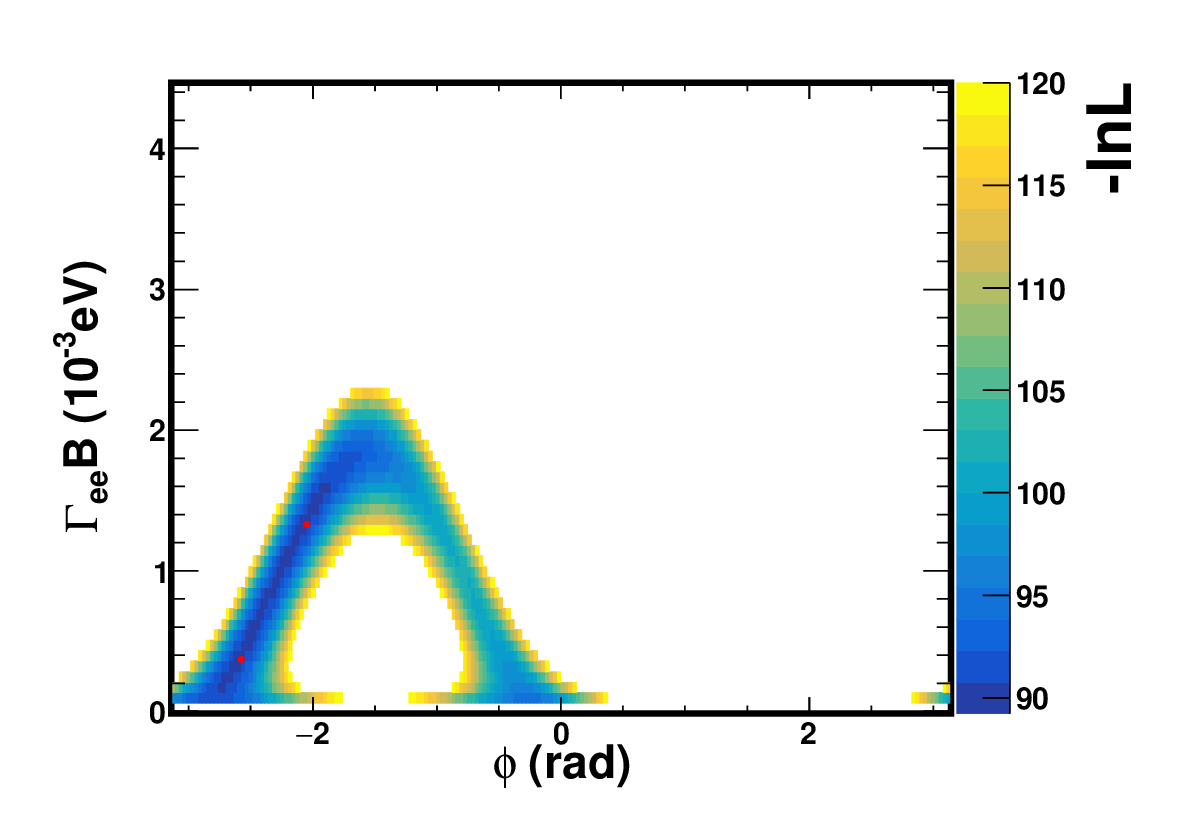}
\end{overpic}

\caption{Distribution of -ln$\mathcal{L}$ as a function of the production  branching fraction multiplied by the electron partial width and the relative phase $\phi$ for the process $\psi(3770)$ $\to$ $\Xi(1530)^{0} \bar{\Xi}^{0}$. The red points represent the nominal values from the best fit. }

\label{fig:chisq}
\end{center}
\end{figure*}

\newpage
%\section{The systematic uncertainty table }
\begin{sidewaystable}[htbp]
\centering
\caption{Total systematic uncertainties~(\%) and their sources for each energy point on BCS measurement. Here, the combined uncertainty from each item is determined with the assumption that the ratio between the signal yield from tagging $\Xi(1530)^0$ and tagging $\bar\Xi(1530)^0$ is $1:1$.} \label{table:sys uncert}
\begin{center}\small
\begin{tabular}{ccccccccccccccccccccccccccccccccccccccccccccc}
\hline\hline
$\sqrt{s}$ (GeV) &3510 &3554 &3582 &3650 &3670 &3682 &3768 &3773 &3780 &3867 &3871 &4040 &4130&4160 &4180 &4190 &4200&4210 &4220 &4230 &4237 &4246 &4260 &4270 &4280 &4290 &4315   \\
\hline
Luminosity &1.0 &1.0 &1.0 &1.0 &1.0 &1.0 &1.0 &0.7 &0.7 &0.7 &0.7&0.7 &0.7&0.7 &0.7&0.7 &0.7&0.7 &0.7&0.7 &0.7 &0.7 &0.7 &0.7 &0.7 &0.7 &0.7  \\
\hline
Tracking  & $1.0$ &1.0 &1.0 &1.0 &1.0 &1.0 &1.0 &1.0 &1.0 &1.0 &1.0&1.0 &1.0 &1.0 &1.0&1.0&1.0 &1.0  &1.0 &$1.0$ &1.0 &1.0 &1.0 &1.0 &1.0 &1.0 &1.0 \\
\hline
PID  & $1.0$ &1.0 &1.0 &1.0 &1.0 &1.0 &1.0 &1.0 &1.0 &1.0 &1.0 &1.0 &1.0&1.0&1.0 &1.0&1.0 & $1.0$ &1.0 &1.0 &1.0 &1.0 &1.0 &1.0 &1.0 &1.0 &1.0\\
\hline
$\Xi^{-}$ Reconstruction  & $5.1$ &5.1 &5.1 &5.1 &5.1 &5.1 &5.1 &5.1 &5.1&5.1 &5.1 &5.1 &5.1&5.1 &5.1 &5.1 &5.1 &5.1 &5.1 & $5.1$ &5.1 &5.1 &5.1 &5.1 &5.1 &5.1 &5.1\\
\hline
$\Lambda $ mass window  & $0.3$ & $0.3$ &0.3 &0.3 &0.3 &0.3 &0.3 &0.3 &0.3 &0.3&0.3 &0.3 &0.3 &0.3 &0.3&0.3 &0.3 &0.3 &0.3 & $0.3$ & $0.3$ &0.3 &0.3 &0.3 &0.3 &0.3 &0.3  \\
\hline
$\Lambda $ decay length  & $0.1$  & $0.1$ &0.1 &0.1 &0.1 &0.1 &0.1 &0.1 &0.1 &0.1&0.1 &0.1 &0.1 &0.1&0.1&0.1 &0.1 &0.1 & $0.1$  & $0.1$ &0.1 &0.1 &0.1 &0.1 &0.1 &0.1  &0.1 \\
\hline
$\Xi^{-}$ mass window  & $1.3$ &1.3 &1.3 &1.3 &1.3 &1.3 &1.3 &1.3 &1.3 &1.3 &1.3&1.3 &1.3 &1.3 &1.3&1.3&1.3 &1.3 & $1.3$ &1.3 &1.3 &1.3 &1.3 &1.3 &1.3 &1.3 &1.3 \\
\hline
$\Xi^{-}$ decay length & $0.5$  &0.5 &0.5 &0.5 &0.5 &0.5 &0.5 &0.5 &0.5 &0.5 &0.5&0.5 &0.5 &0.5 &0.5&0.5&0.5 &0.5 & $0.5$  &0.5 &0.5 &0.5 &0.5 &0.5 &0.5 &0.5 &0.5  \\
\hline
$\Xi(1530)^{0}$ mass window  &$3.6$ &3.6  &3.6 &3.6 &$-$ &3.6 &3.6 &3.6 &3.6 &3.6 &3.6&3.6 &3.6 &3.6 &3.6&3.6&3.6 &3.6 &$3.6$ &3.6  &3.6 &3.6 &3.6 &3.6 &3.6 &3.6  &3.6 \\
\hline
$\bar{\Xi}^{0}$ mass window  &$2.3$ &2.3 &2.3 &2.3 &2.3 &2.3 &2.3 &2.3 &2.3 &2.3 &2.3&2.3 &2.3 &2.3 &2.3&2.3 &2.3&2.3 &$2.3$ &2.3 &2.3 &2.3 &2.3 &2.3 &2.3 &2.3  &2.3 \\
\hline
Scale factors   &$1.1$ &1.1 &1.1 &1.1 &1.1 &1.1 &1.1 &1.1 &1.1 &1.1 &1.1 &1.1 &1.1&$1.1$ &1.1 &1.1 &1.1&$1.1$&1.1 &1.1 &1.1 &1.1 &1.1 &1.1 &1.1 &1.1 &1.1  \\
\hline
Wrong combinations  & 2.1 & 2.1  & 2.1 & 2.1 & 2.1 & 2.1 & 2.1 & 2.1 & 2.1 & 2.1 & 2.1 & 2.1 & 2.1 & 2.1 & 2.1 & 2.1 & 2.1 & 2.1 & 2.1 & 2.1 & 2.1 & 2.1 & 2.1 & 2.1 & 2.1 & 2.1 & 2.1  \\
\hline
Angular distribution  & $4.3$ &4.3 &4.3 &4.3 &4.3 &4.3 &4.3  &4.3 &4.3 &4.3 &4.3&4.3 &4.3 &4.3 &4.3 &4.3&4.3 &4.3 & $4.3$ &4.3 &4.3 &4.3 &4.3 &4.3 &4.3 &4.3 &4.3  \\
\hline
Input lineshape & 3.9  &3.9 &3.9 &3.9 &3.9 &3.9 &3.9 &3.9 &3.9 &3.9 &3.9&3.9 &3.9 & 3.9  &3.9&3.9 &3.9 & 3.9&3.9 &3.9 &3.9 &3.9 &3.9&3.9 &3.9 &3.9 &3.9  \\
\hline
$Br(\Xi(1530)^{0} \to \Xi^{-} \pi^{+})$ & $3.7$ &3.7  &3.7 &3.7 &3.7 &3.7 &3.7&3.7 &3.7 &3.7 &3.7&3.7 &3.7&3.7 &3.7 &3.7 &3.7 &3.7 & $3.7$ &3.7 &3.7 &3.7 &3.7 &3.7 &3.7 &3.7 &3.7  \\
\hline
$Br(\Xi^{-} \to \Lambda \pi^{-})$  & $0.0$ &0.0 &0.0 &0.0 &0.0 &0.0 &0.0 &0.0&0.0 &0.0 &0.0 &0.0 &0.0 &0.0 & $0.0$ &0.0  &0.0 &0.0&0.0 &0.0 &0.0 &0.0 &0.0 &0.0&0.0 &0.0 &0.0 \\
\hline
$Br(\Lambda \to p\pi^{-})$  & $0.8$ &0.8  &0.8 &0.8 &0.8 &0.8 &0.8&0.8 &0.8 &0.8 &0.8 &0.8 &0.8&0.8 &0.8 &0.8 &0.8 &0.8 & $0.8$ &0.8  &0.8 &0.8 &0.8 &0.8 &0.8 &0.8 &0.8  \\
\hline

Total  &10.1&10.1 &10.1  &10.1 &10.1 &10.1&10.1&10.1 &10.1 &10.1&10.1&10.1&10.1&10.1&10.1&10.1&10.1&10.1&10.1&10.1&10.1 &10.1&10.1&10.1&10.1&10.1&10.1 \\
\hline
\hline
\end{tabular}
\end{center}
\end{sidewaystable}

\newpage
%\section{The systematic uncertainty table }
\begin{sidewaystable}[htbp]
\centering
\caption{Total systematic uncertainties~(\%) and their sources for each energy point on BCS measurement. Here, the combined uncertainty from each item is determined with the assumption that the ratio between the signal yield from tagging $\Xi(1530)^0$ and tagging $\bar\Xi(1530)^0$ is $1:1$.} \label{table:sys uncert}
\begin{center}\small
\begin{tabular}{cccccccccccccccccccccc}
\hline\hline
$\sqrt{s}$ (GeV) &4340&4360&4380&4400&4420&4440&4470 &4530 &4600 &4612&4620 &4640 &4660&4680&4700&4740&4750&4780&4840&4914&4946 \\
\hline
Luminosity  &0.7&0.7 &0.7&0.7&0.7&0.7 &0.7&0.7&0.5&0.5&0.5 &0.5 &0.5 &0.5&0.5&0.5 &0.5 &0.5 &0.5&0.5 &0.5\\
\hline
Tracking   &1.0 &1.0 &1.0 &1.0 &1.0 &1.0 &1.0&1.0 &1.0 &1.0&1.0 &1.0&1.0 &1.0 &1.0&1.0&1.0 &1.0&1.0 &1.0 &1.0\\
\hline
PID   &1.0 &1.0 &1.0 &1.0 &1.0 &1.0&1.0 &1.0 & $1.0$ &1.0&1.0 &1.0&1.0 &1.0 &1.0&1.0&1.0 &1.0&1.0 &1.0 &1.0\\
\hline
$\Xi^{-}$ Reconstruction &5.1 &5.1 &5.1 &5.1&5.1 &5.1 &5.1&5.1 &5.1 &5.1&5.1 &5.1&5.1 &5.1 &5.1&5.1&5.1 &5.1&5.1 &5.1 &5.1\\
\hline
$\Lambda $ mass window  &0.3 &0.3 &0.3 &0.3  &0.3&0.3 &0.3&0.3 &0.3 &0.3&0.3 &0.3&0.3 &0.3 &0.3&0.3&0.3 &0.3&0.3 &0.3 &0.3\\
\hline
$\Lambda $ decay length   &0.1 &0.1 &0.1 &0.1 &0.1&0.1 &0.1&0.1 &0.1 &0.1&0.1 &0.1&0.1 &0.1 &0.1&0.1&0.1 &0.1&0.1 &0.1 &0.1\\
\hline
$\Xi^{-}$ mass window   &1.3 &1.3 &1.3 &1.3 &1.3&1.3 &1.3&1.3 &1.3 &1.3&1.3 &1.3&1.3 &1.3 &1.3&1.3&1.3 &1.3&1.3 &1.3 &1.3\\
\hline
$\Xi^{-}$ decay length  &0.5 &0.5 &0.5 &0.5 &0.5 &0.5&0.5&0.5 &0.5 &0.5&0.5&0.5&0.5 &0.5 &0.5&0.5&0.5&0.5&0.5 &0.5 &0.5\\
\hline
$\Xi(1530)^{0}$ mass window &3.6 &3.6 &3.6 &3.6 &3.6&3.6 &3.6&3.6 &3.6 &3.6&3.6 &3.6&3.6 &3.6 &3.6&3.6&3.6 &3.6&3.6 &3.6 &3.6\\
\hline
$\bar{\Xi}^{0}$ mass window   &2.3 &2.3 &2.3 &2.3 &2.3&2.3 &2.3&2.3 &2.3 &2.3&2.3 &2.3&2.3 &2.3 &2.3&2.3&2.3 &2.3&2.3 &2.3 &2.3\\
\hline
Scale factors    &1.1 &1.1 &1.1 &1.1 &1.1 &1.1&1.1&1.1 &1.1 &1.1&1.1&1.1&1.1 &1.1 &1.1&1.1&1.1&1.1&1.1 &1.1 &1.1\\
\hline
Wrong combinations   & 2.1 & 2.1 & 2.1 & 2.1 & 2.1 & 2.1& 2.1 & 2.1 & 2.1 & 2.1 & 2.1 & 2.1& 2.1 & 2.1 & 2.1& 2.1 & 2.1 & 2.1& 2.1 & 2.1 & 2.1\\
\hline
Angular distribution  &4.3 &4.3 &4.3 &4.3 &4.3 &4.3&4.3&4.3 &4.3 &4.3&4.3&4.3&4.3 &4.3 &4.3&4.3&4.3&4.3&4.3 &4.3 &4.3\\
\hline
Input lineshape & 3.9  &3.9 &3.9 &3.9 &3.9 &3.9 &3.9 &3.9 &3.9 &3.9 &3.9&3.9 &3.9 & 3.9  &3.9&3.9 &3.9 & 3.9&3.9 &3.9 &3.9\\
\hline
$Br(\Xi(1530)^{0} \to \Xi^{-} \pi^{+})$ &3.7 &3.7 &3.7&3.7 &3.7 &3.7 &3.7&3.7 &3.7 &3.7 &3.7 &3.7&3.7 &3.7&3.7&3.7 &3.7 &3.7&3.7 &3.7&3.7\\
\hline
$Br(\Xi^{-} \to \Lambda \pi^{-})$  &0.0 &0.0 &0.0 &0.0 &0.0&0.0 &0.0&0.0 &0.0 &0.0&0.0 &0.0&0.0 &0.0&0.0&0.0&0.0 &0.0&0.0 &0.0&0.0\\
\hline
$Br(\Lambda \to p\pi^{-})$    &0.8 &0.8 &0.8 &0.8 &0.8&0.8 &0.8&0.8 &0.8 &0.8&0.8&0.8 &0.8&0.8 &0.8&0.8&0.8&0.8 &0.8&0.8 &0.8\\
\hline

Total   &10.1&10.1 &10.1  &10.1 &10.1 &10.1&10.1&10.1 &10.1 &10.1&10.1&10.1&10.1&10.1&10.1&10.1&10.1&10.1&10.1&10.1&10.1\\
\hline
\hline
\end{tabular}
\end{center}
\end{sidewaystable}

\FloatBarrier
\bibliography{basename of .bib file}

%\end{thebibliography}